\pdfoutput=1

\documentclass[final,prl,aps,twocolumn,preprintnumbers,amssymb,nobibnotes,nofootinbib,superscriptaddress,10pt]{revtex4-1}
\usepackage{hyperref,graphicx,array,multirow,color,amsmath,mathrsfs,titlesec,shuffle}

\usepackage{slashed}
\titleformat{\section}{\centering\normalsize\normalfont\bf}{\thesection}{1em}{}

\hypersetup{pdftitle={},pdfcreator={},linkcolor=[rgb]{0.15,0.35,0.75},colorlinks=true,citecolor=[rgb]{0.675,0,0.2},urlcolor=[rgb]{0.15,0.35,0.65}}

\newcommand{\fwbox}[2]{\text{\makebox[#1][c]{$\hspace{-150pt}\displaystyle#2\hspace{-150pt}$}}}
\newcommand{\fwboxL}[2]{\text{\makebox[#1][l]{$#2$}}}
\newcommand{\fwboxR}[2]{\text{\makebox[#1][r]{$#2$}}}
\renewcommand{\phi}{\varphi}

\newcommand{\eq}[1]{\vspace{-3.5pt}\begin{equation}\hspace{2pt}#1\hspace{-0pt}\vspace{-3.5pt}\end{equation}}

\newcommand{\fig}[2]{\vcenter{\includegraphics[scale=#1]{./figures/#2}}}

\newcommand{\ab}[1]{\langle #1 \rangle}
\newcommand{\sqb}[1]{[ #1 ]}
\newcommand{\sab}[1]{s_{#1}}

\newcommand{\mi}{\raisebox{0.75pt}{\scalebox{0.75}{$\hspace{-1pt}\,-\,\hspace{-0.75pt}$}}}
\renewcommand{\pl}{\raisebox{0.75pt}{\scalebox{0.75}{$\hspace{-1pt}\,+\,\hspace{-0.75pt}$}}}

\def\lra{\leftrightarrow}

\def\A{\mathcal{A}}
\def\O{\mathcal{O}}
\def\N{\mathcal{N}}

\def\I{\mathcal{I}}

\def\PT{\text{PT}}
\def\tr{\text{tr}}

\def\eps{\epsilon}

\newcommand{\AST}[1]{A^{\text{ST}}[#1]}
\newcommand{\ADT}[1]{A^{\text{DT}}[#1]}
\newcommand{\ASLST}[1]{A^{\text{SLST}}[#1]}

\newcommand{\NeqFour}{\mathcal{N}\!=\!4\text{ sYM}}

\definecolor{airforceblue}{rgb}{0.36, 0.54, 0.66}
\definecolor{bananayellow}{rgb}{1.0, 0.88, 0.21}
\definecolor{bittersweet}{rgb}{1.0, 0.44, 0.37}
\definecolor{blue(ncs)}{rgb}{0.0, 0.53, 0.74}
\definecolor{bole}{rgb}{0.47, 0.27, 0.23}
\definecolor{brass}{rgb}{0.71, 0.65, 0.26}
\definecolor{bronze}{rgb}{0.8, 0.5, 0.2}
\definecolor{brgreen}{rgb}{0.0, 0.26, 0.15}
\definecolor{burgundy}{rgb}{0.5, 0.0, 0.13}
\definecolor{cherry}{rgb}{1.0, 0.72, 0.77}
\definecolor{cocao}{rgb}{0.82, 0.41, 0.12}
\definecolor{citrine}{rgb}{0.99, 0.82, 0.07}

\makeatletter\DeclareRobustCommand*{\bfseries}{\not@math@alphabet\bfseries\mathbf\fontseries\bfdefault\selectfont\boldmath}\makeatother

\begin{document}

\title{The two-loop five-point amplitude in $\N=4$ super-Yang--Mills theory}
\preprint{CP3-18-81, IPhT-18/170, SLAC--PUB--17369}


%
\author{Samuel~Abreu}
\affiliation{Center for Cosmology, Particle Physics and Phenomenology (CP3),
Universit\'e Catholique de Louvain, 1348 Louvain-La-Neuve, Belgium}
\author{Lance~J.~Dixon}
\affiliation{SLAC National Accelerator Laboratory, Stanford University, Stanford, CA 94039, USA}
\author{Enrico~Herrmann}
\affiliation{SLAC National Accelerator Laboratory, Stanford University, Stanford, CA 94039, USA}
\author{Ben~Page}
\affiliation{Institut de Physique Th\'eorique, CEA, CNRS, Universit\'e Paris-Saclay, F-91191 Gif-sur-Yvette cedex, France}
\author{Mao~Zeng}
\affiliation{Institut f\"ur Theoretische Physik, Eidgen\"ossische Technische Hochschule Z\"urich,
Wolfgang-Pauli-Strasse 27, 8093 Z\"urich, Switzerland}

\begin{abstract}
We compute the symbol of the
two-loop five-point scattering amplitude in $\mathcal{N}$ = 4
super-Yang--Mills theory, including its full color dependence.
This requires constructing the symbol of all two-loop five-point 
nonplanar massless master integrals, for which we give explicit 
results.
\end{abstract}
\maketitle
A great deal of progress in calculating scattering amplitudes has been 
driven by the fruitful interplay between new formal ideas 
and the need for increasingly precise theoretical predictions
at collider experiments. For instance,
techniques such as generalized unitarity~\cite{Bern:1994zx} and
the symbol calculus~\cite{Goncharov:2010jf} 
were first introduced in the realm of 
maximally supersymmetric Yang--Mills theory ($\NeqFour$) and 
went on to have a large impact on precision collider physics. In this letter, we use cutting-edge techniques to take a first look at the analytic 
form of the two-loop five-point amplitude in $\NeqFour$ beyond 
the planar, $N_c\!\to\!\infty$, limit of $SU(N_c)$ gauge theory.

Amplitudes in $\NeqFour$ possess rigid analytic properties that 
make them easier to compute than their pure Yang--Mills
counterparts, the state of the art being the
three-loop four-gluon $\NeqFour$ amplitude \cite{Henn:2016jdu}.
Historically, calculations in $\NeqFour$ have therefore preceded
analogous computations in QCD.  The planar five-point amplitude at
two loops in $\NeqFour$ was first obtained
numerically~\cite{Bern:2006vw}, confirming the prediction
of \cite{BDS}. In pure Yang--Mills, the first planar two-loop
five-point amplitude, evaluated numerically,
was for the all-plus helicity configuration~\cite{Badger:2013gxa}. Since then, a flurry of activity in \emph{planar}
multi-leg two-loop amplitudes has seen the analytic calculation
of the all-plus amplitude~\cite{Gehrmann:2015bfy}, the numerical
evaluation of all five-parton QCD amplitudes~\cite{Badger:2017jhb, Badger:2018gip, Abreu:2017hqn,
Abreu:2018jgq}, and recently the computation of 
analytic expressions for all five-gluon scattering amplitudes~\cite{Badger:2018enw,Abreu:2018zmy}. These achievements were 
made possible by the development of efficient ways to
reduce amplitudes to master integrals using integration-by-parts (IBP) relations~\cite{IBP1,IBP2},
automated by Laporta's algorithm~\cite{Laporta:2001dd}, or modern reformulations based on unitarity cuts and computational algebraic geometry~\cite{Gluza:2010ws,Ita:2015tya, Larsen:2015ped, Boehm:2018fpv,Larsen:2015ped,Abreu:2017hqn}, and to compute master integrals from their differential equations~\cite{Gehrmann:1999as,Henn}. Indeed, all planar five-point
master integrals have now been computed~\cite{Papadopoulos:2015jft,Gehrmann:2018yef}, 
and substantial progress has been made in the nonplanar
sectors as well~\cite{Chicherin:2017dob,Abreu:2018rcw,Chicherin:2018mue}.

In this work, we first discuss the \emph{integrand} of the 
two-loop five-point amplitude in $\NeqFour$, and how it can be 
reduced to a form 
involving only so-called pure integrals (i.e., integrals 
satisfying a differential equation in canonical 
form~\cite{Henn}).
We then use the aforementioned new techniques for integral 
reduction and
differential equations (most notably the method introduced in 
\cite{Abreu:2018rcw}) to compute the \emph{symbols} 
\cite{Goncharov:2010jf} 
(see also \cite{Duhr:2011zq,Duhr:2012fh}) of
\emph{all} nonplanar massless two-loop five-point master integrals. 
From these integrals we finally assemble the symbol of the complete two-loop five-point $\NeqFour$ amplitude and discuss consistency checks of
our result. Throughout, we work at the level of the symbol where
transcendental constants are set to zero.
While such contributions are important for the numerical evaluation of an
amplitude, the symbol itself contains a major part of the non-trivial analytic structure of the amplitude.

Our result constitutes the first analytic investigation of two-loop five-point amplitudes in any gauge or gravity theory beyond the planar limit.
Just as the one-loop five-gluon amplitude \cite{Bern:1993mq} did, our two-loop result should provide valuable theoretical data
for further exploring the properties of structurally complex
amplitudes, as well as the proposed duality
between scattering amplitudes and Wilson loops at subleading 
color \cite{Ben-Israel:2018ckc}. Furthermore, the methods will 
impact precision collider phenomenology: the master integrals are
directly applicable to QCD amplitudes, opening the way to
computing three-jet production at hadron colliders at next-to-next-to-leading order.

\vspace{-8pt}
\section{Construction of the amplitude} 
\vspace{-6pt}
\label{sec:amplitude_decomposition}
%
In any $SU(N_c)$ gauge theory with all states in the adjoint 
representation, the trace-based color decomposition \cite{Dixon:1996wi,BRY} of any two-loop five-point amplitude is\footnote{\label{footn:norm}In our conventions, we factor out $\frac{e^{-\gamma\eps} g^2 N_c}{(4\pi)^{2-\eps}}$ per loop in a standard perturbative expansion of the amplitude. 
The generators of the fundamental representation of $SU(N_c)$ are normalized as $\tr[T^aT^b] = \delta^{ab}$, and $\tr[i_1i_2...i_k] \equiv \tr[T^{a_{i_1}}T^{a_{i_2}}\cdots T^{a_{i_k}}]$.}
\begin{align}
\vspace{-.2cm}
	\label{eq:trace_decomp}
	&\!\A^{(2)}_5=\!\!\!\!\!\!\!\!\!\!\sum_{S_5/(S_3\times Z_2)}
	\!\!\!\!\!\!\!\!
	\frac{\tr [15]\big(\tr[234]\mi\tr[432]\big)}{N_c} 
	 \ADT{15|234}\\[-4pt]
	&\!\!\!+\!\!\!\!\sum_{S_5/D_5 } \!\!\!\!
	\big(\tr [12345]\mi \tr [54321]\big)
	\! \bigg(\!\! \AST{12345}
		\pl \!\frac{\ASLST{12345}}{N^2_c} \! \bigg).\hspace{-.2cm} \nonumber \vspace{-1cm}
\end{align}
\vskip-.2cm
Here, single-trace (ST), subleading-color single-trace (SLST)
and double-trace (DT) denote different 
\emph{partial amplitudes}.
$S_n$ ($Z_n$) is
the (cyclic) permutation group, and $D_n$ the dihedral group.

It is a powerful fact about MHV scattering
amplitudes in $\NeqFour$ that all leading 
singularities~\cite{Cachazo:2008vp} are given 
in terms of different permutations of 
Parke-Taylor tree-(super-)amplitudes~\cite{Parke:1986gb,Nair:1988bq}. 
This highly nontrivial result has been derived from a dual 
formulation of leading singularities in terms of the 
Grassmannian~\cite{Arkani-Hamed:2014bca}. Furthermore, 
$\NeqFour$ amplitudes are conjectured to be of uniform
transcendental weight~\cite{Dixon:2011pw,BDS,ArkaniHamed:2012nw,LipatovTranscendentality}.
A representation of the \emph{four}-dimensional integrand
has been given in~\cite{Bern:2015ple},
where this Parke-Taylor structure, together with further special analytic properties of $\NeqFour$ (logarithmic singularities and no 
residues at infinite loop momentum), is manifest.
In this representation, the full, color-dressed amplitude splits into 
three distinct parts
\eq{
\A^{(2)}_n = C \otimes \PT \otimes g_{\text{pure}}\,, 
\label{eq:schematic_amp_color_pt_split}
}
where $C$ schematically denotes the color structure of the gauge
theory. For a five-point scattering amplitude, the space of
Parke-Taylor factors 
is spanned by a set of $3!$ Kleiss-Kuijf (KK) 
independent elements \cite{Kleiss:1988ne} that we denote by 
$\PT[1\sigma_2\sigma_3\sigma_45]$, where
\eq{
\label{eq:pt_def}
\hspace{.1cm}
\PT[\sigma_1\sigma_2\sigma_3\sigma_4\sigma_5] \!=\! \frac{\delta^8(Q)}{\ab{\sigma_1\sigma_2}\ab{\sigma_2\sigma_3}\ab{\sigma_3\sigma_4}\ab{\sigma_4\sigma_5}\ab{\sigma_5\sigma_1}}.
\hspace{-.4cm}
}
The super-momentum conserving delta-function, $\delta^8(Q)$, encodes the supersymmetric Ward-identities relating the $(\mi\mi\pl\pl\pl)$-helicity five-gluon amplitude to all other five-particle amplitudes. 
The third part, $g_{\text{pure}}$, denotes a pure function of
transcendental weight $4$.

\begin{figure}[]
\vspace{-14pt}
$
\fwboxR{24pt}{\fig{.30}{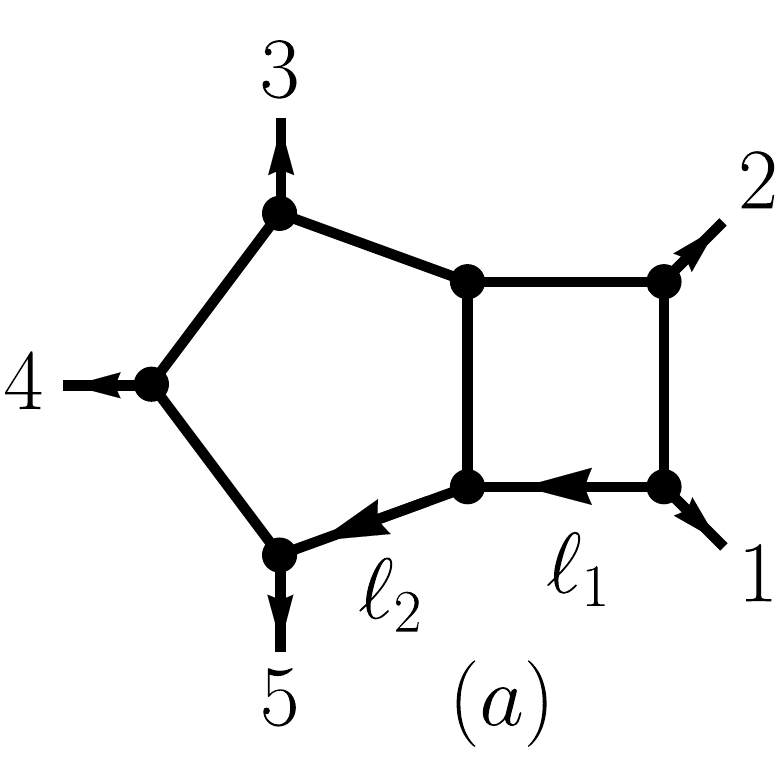}} 
\vspace{-7pt} 
\fwbox{-75pt}{\fig{.30}{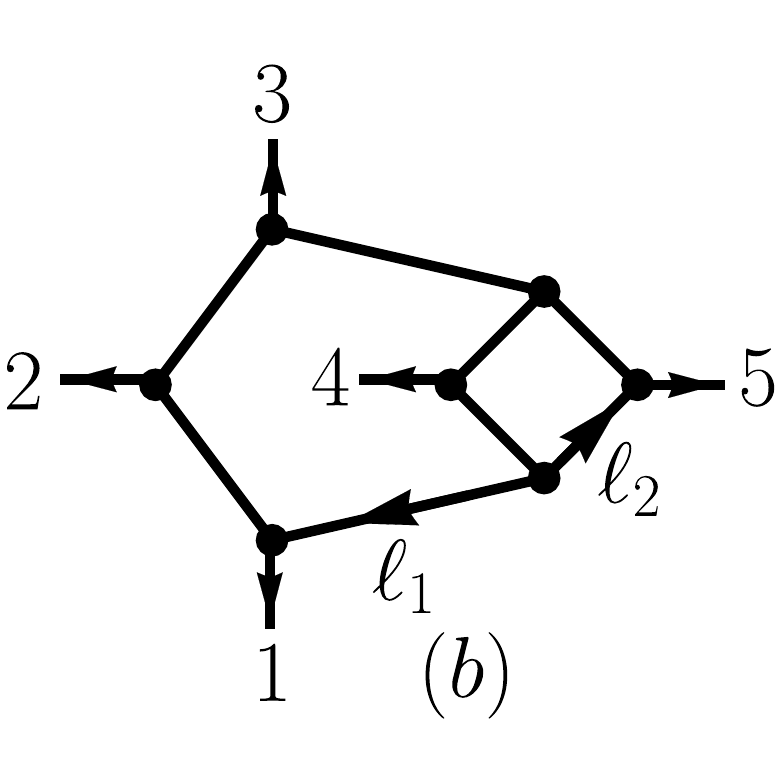}} 
\raisebox{-2pt}{\fwboxL{24pt}{\fig{.30}{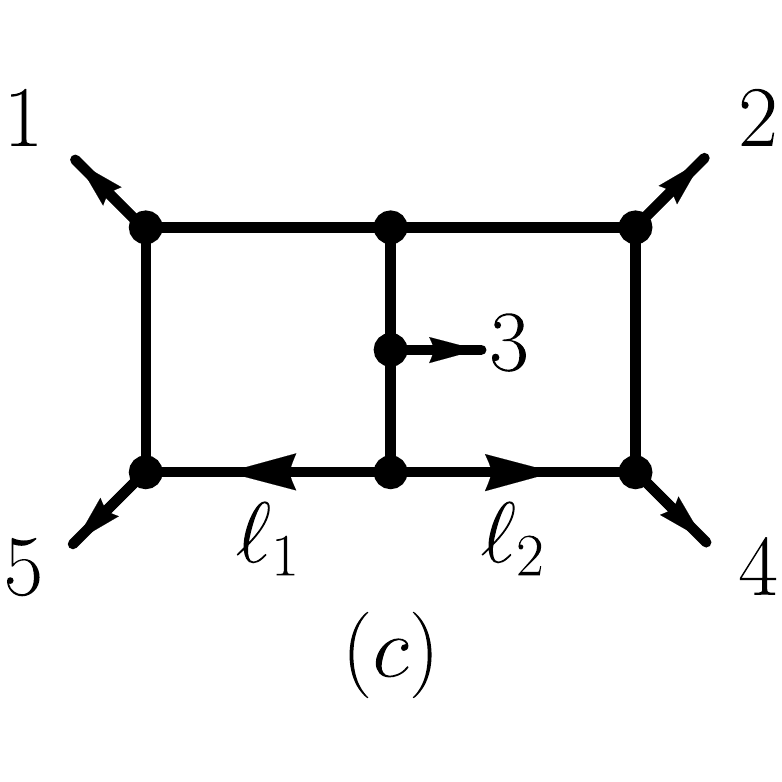}}} $ 
\\
$
\fwboxR{24pt}{\fig{.30}{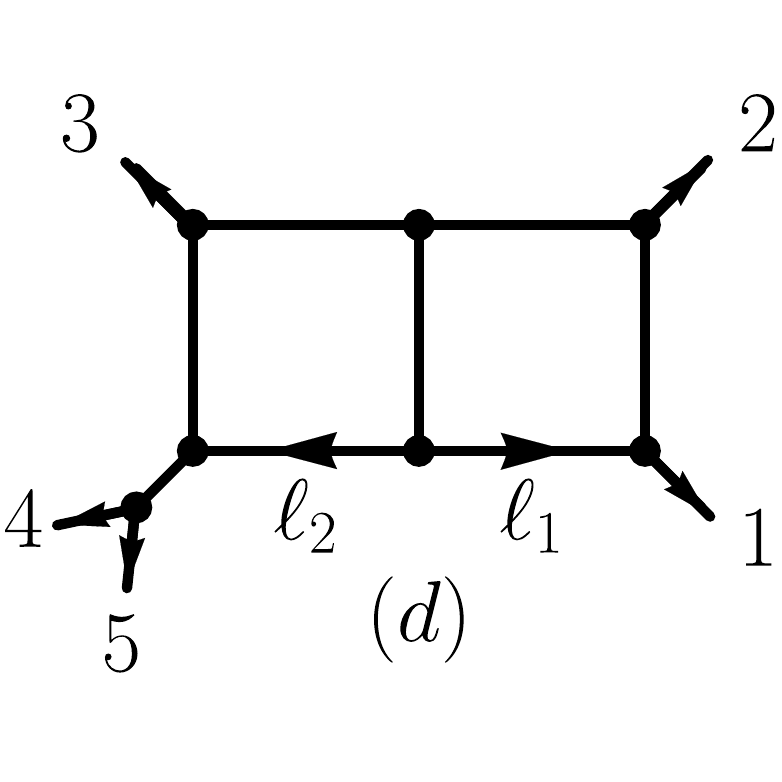}} 
\fwbox{-78pt}{\fig{.30}{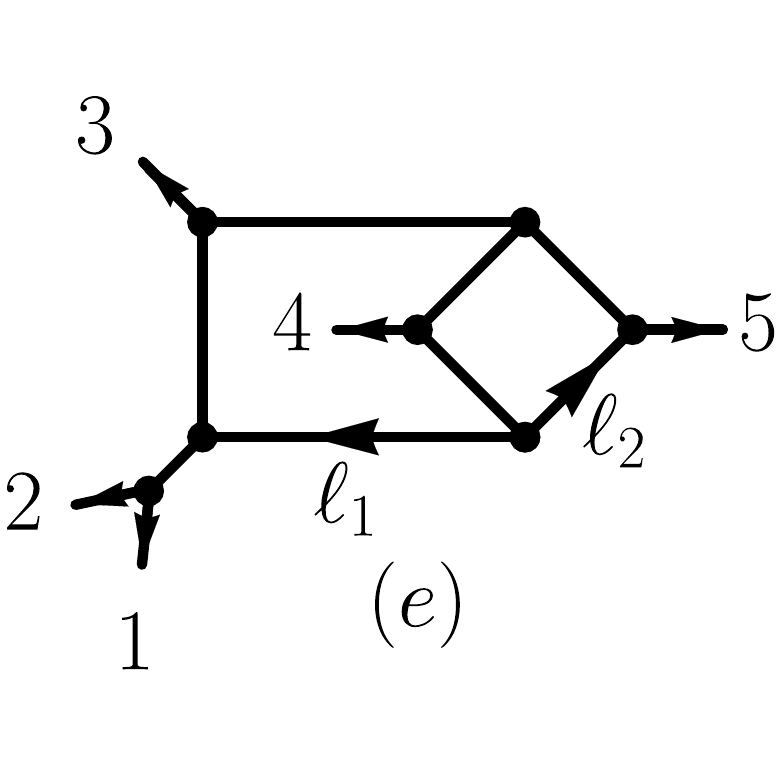}} 
\raisebox{3pt}{
\fwboxL{24pt}{\fig{.30}{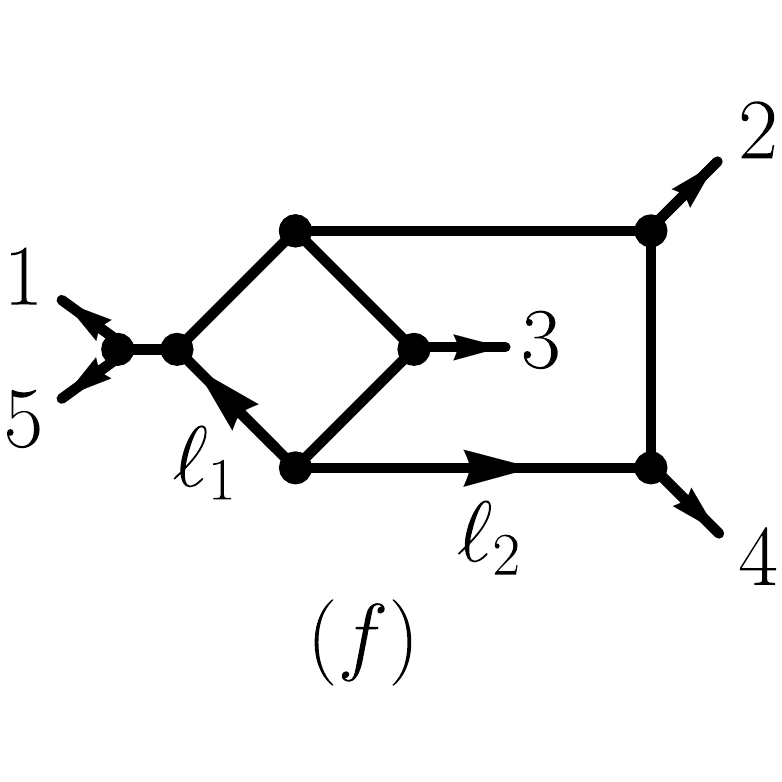}} }
$ 
\vspace{-7pt}
\caption{\label{fig:bcj_integrands}Diagram topologies entering the local representation of the two-loop five-point integrand of $\NeqFour$ \cite{Carrasco:2011mn}. Each diagram has an associated color structure and numerator which we suppress.}
\end{figure}

The goal of this section is to compute the partial amplitudes 
in~\eqref{eq:trace_decomp}. Our starting point is the integrand of~\cite{Carrasco:2011mn} which
is valid in $d=4\mi2\epsilon$ space-time dimensions 
and is given in terms of the six topologies in Fig.~\!\ref{fig:bcj_integrands},
\eq{
\hspace{.3cm}
\A^{(2)}_5 =  \sum_{S_5} \left(\frac{ I^{(a)}}{2} \pl \frac{I^{(b)}}{4} \pl \frac{ I^{(c)}}{4} \pl \frac{I^{(d)} }{2} \pl \frac{ I^{(e)}}{4} \pl \frac{I^{(f)}}{4} \right)\,. 
\hspace{-.3cm}
\label{eq:amp_bcj_rep}
}
The sum is over all $5!$ permutations of external legs
and the rational numbers correspond to diagram-symmetry factors.

For each of the topologies in Fig.~\!\ref{fig:bcj_integrands}, we construct a basis of pure master integrals, on which the amplitude \eqref{eq:amp_bcj_rep} can be
decomposed, so that the separation into color, rational, and transcendental parts \eqref{eq:schematic_amp_color_pt_split} becomes manifest. 
Most required master integrals are already known
in pure form \cite{Gehrmann:2015bfy,Papadopoulos:2015jft, 
Gehrmann:2001ck, Chicherin:2018mue, Abreu:2018rcw}. 
The one missing topology, which we discuss momentarily, is the nonplanar
double-pentagon (diagram~(c) of Fig.~\ref{fig:bcj_integrands}).
The integrals we are concerned with are functions 
of five Mandelstam invariants, $s_{12}, s_{23}, s_{34}, s_{45}, s_{51}$, with $s_{ij}\! =\! (k_i \pl k_j)^2$. 
We also encounter the parity-odd $\varepsilon$-tensor
contraction
\begin{equation}
\tr_5 = 4 i \varepsilon _ { \mu \nu \rho \sigma } k _ { 1 } ^ 
{\mu } k _ { 2 } ^ { \nu } k _ { 3 } ^ { \rho } k _ { 4 } ^ { \sigma } = \tr (\gamma^5 \slashed k_1 \slashed k_2 \slashed k_3 \slashed k_4) \, .
\label{eq:tr5Def}
\end{equation}
To find a basis of pure master integrals for the top-level (eight-propagator) topology of Fig.~\ref{fig:bcj_integrands}(c) it was necessary
to construct nine independent numerators. Specifically, we chose the following set of master integrals:
\begin{enumerate}
	\item The parity-even part of the integral with numerator $N_1^{(a)}$
	identified in~\cite{Bern:2015ple}, rewritten as spinor 
	traces in Eq.~(21) of~\cite{Bern:2018oao}. By deleting 
	$\gamma^5$ from the spinor traces, we obtain the parity-even 
	parts in a form that is valid in $d$ dimensions. Two more 
	pure integrals are obtained from it by using the diagram's $Z_2\times Z_2$ symmetry.
	\item $(6\mi2\epsilon)$-dimensional scalar integrals 
	with any
	of the eight propagators squared, normalized by a
	factor of $\tr_5$ and a homogeneous linear function of the
	$s_{ij}$ variables.
	Six such integrals, which we have converted 
	to integrals in $(4\mi2\epsilon)$ dimensions 
	\cite{Bern:1992em,Bern:1993kr,Tarasov:1996br,
	Lee:2009dh}, are included in our basis.
\end{enumerate}
Explicit expressions for these new pure master integrals can be
found in the ancillary file {\tt masters.m}.

Next, we construct differential equations in canonical 
form~\cite{Henn} for the master integrals. The (iterated)
branch-cut structure of the integrals is encoded 
in the \emph{symbol letters} which are algebraic functions of the kinematic invariants. It is convenient to parametrize the five-point
kinematics in terms of variables that rationalize all letters of the \emph{alphabet}. 
This can be accomplished via 
momentum-twistors~\cite{Hodges} and the $x_i$-parametrization proposed in 
\cite{Badger:2013gxa}.\footnote{\label{footnoteKin}
	Explicit expressions for our kinematics can be found in the ancillary file \texttt{kinematics.m}.} 
For the nonplanar double-pentagon integral, we find that the complete system contains 108 masters and depends on the 31
$W_{\alpha}$-letters suggested in \cite{Chicherin:2017dob}:
\eq{
\hspace{.1cm}
\partial_{x_i}  \I_a \!\equiv\! \frac{\partial  \I_a}{\partial x_i}\! =\! \epsilon \sum_{\alpha=1}^{31}
\frac{\partial \log W_\alpha} {\partial x_i}  M_\alpha^{ab} \,   \I_b \,, \ 1\!\leq\! a, b \!\leq\! 108 \,.  \hspace{-.5cm}
\label{eq:diffCanonical}
}
Ten of the letters ($\alpha\!\!\in\!\! \{1,\!...,\!5\}\!\!\cup\!\!\{16,\!...,\!20\}\!$) are simple Mandelstam invariants $\sab{ij}$, 
15 further letters  ($\alpha\!\in\!\{6,\!...,\!15\}\!\cup\!\{21,\!...,\!25\}$) are differences of Mandelstam invariants $\sab{ij} \mi \sab{kl}$, 
the 5 parity-odd letters ($\alpha \!\!\in\!\!\{26,\!...,30\}$) can be expressed as ratios of spinor-brackets such as
$\frac{\ab{12}\sqb{15}\ab{45}\sqb{24}}{\sqb{12}\ab{15}\sqb{45}\ab{24}}$ which invert under complex conjugation $\ab{\cdot}\!\lra\!\sqb{\cdot} \text{ or } \tr_5\!\to\!\mi\tr_5$, 
and the final, parity-even letter ($\alpha\!=\!31$) is $\tr_5$. The $31$ $M_\alpha$-matrices consist of simple rational numbers. 

Computing the $M_\alpha$-matrices in~\eqref{eq:diffCanonical}
requires performing IBP reduction on differentials of the original masters $\partial_{x_i}\I_a$ with respect to the kinematic variables in order to re-express them in terms of the original basis $\I_a$. We use the 
efficient approach introduced in~\cite{Abreu:2018rcw}, which
builds on the modern formulation of IBP relations in terms of unitarity cuts
and computational algebraic geometry~\cite{Gluza:2010ws,Ita:2015tya, Larsen:2015ped,
Boehm:2018fpv,Larsen:2015ped,Abreu:2017hqn}.
The method requires IBP reduction at only $30$ rational, 
numerical phase-space points to fix all the $M_\alpha$,
dramatically reducing the computation time compared to analytic 
IBP reduction. Combined with the \emph{first-entry condition}~\cite{Gaiotto:2011dt}, which restricts integrals to only
have branch-cut singularities at physical thresholds,
we obtain solutions to the differential equations at the 
symbol level for all master integrals. As a check, we verified
that we reproduce (at symbol level) all known results for 
descendant integrals ($\leq 7$ propagators).
The full results are included in the ancillary file 
\texttt{masters.m}.

Having established a basis and computed the master integrals
required for massless two-loop five-point amplitudes, we can now write the $\NeqFour$ amplitude in that basis.
As already stated, we use the $d$-dimensional 
representation of the integrand given in \cite{Carrasco:2011mn}.
While this representation has the
advantage of being in the so-called 
Bern-Carrasco-Johansson (BCJ) form~\cite{BCJLoop}, 
which allows for the immediate construction of the gravity 
integrand via the `double-copy' prescription, it obscures some
of the simplicity of the final result. For instance, each
individual diagram in Fig.~\!\ref{fig:bcj_integrands} introduces
spurious rational factors. Applying Fierz 
color-identities~\cite{Dixon:1996wi} to decompose the 
integrand \eqref{eq:amp_bcj_rep} into the partial amplitudes in 
\eqref{eq:trace_decomp} and using IBP reduction to
rewrite those in our pure basis, we can obtain a representation
that is manifestly in the form of 
\eqref{eq:schematic_amp_color_pt_split}. 
In particular, we find a simple rational kinematic dependence for all partial amplitudes via at most six KK-independent Parke-Taylor factors:
\begin{align}
\label{eq:structural_result_partial_amps_pt_decomp}
\begin{split}
	\AST{12345}  & = \PT[12345] \ \text{M}^{\text{BDS}}_{(2)} \,,   \\[1pt]
	\ADT{15|234} & =\!\!\sum_{\sigma(234) \in S_3} \hspace{-.4cm} \PT[1\sigma_2\sigma_3\sigma_45]\  g^{\text{DT}}_{\sigma_2\sigma_3\sigma_4}\,, \\[-2pt]
	\ASLST{12345} & =\!\!\sum_{\sigma(234)\in S_3}\hspace{-.4cm} \PT[1\sigma_2\sigma_3\sigma_45]\ g^{\text{SLST}}_{\sigma_2\sigma_3\sigma_4}\,, 
\end{split}
\end{align}
\vskip -.3cm \noindent
where $\text{M}^{\text{BDS}}_{(2)}$ is the two-loop
BDS ansatz~\cite{BDS} and $g^\text{X}_{\vec{\sigma}}$ are pure
functions.  
Both $\text{M}^{\text{BDS}}_{(2)}$ and $g^\text{X}_{\vec{\sigma}}$
can be written as $\mathbb{Q}$-linear combinations of our pure
master integrals.
The IBP reduction was done following the same strategy already
discussed for the differential equations. Given the simple
kinematic dependence of the result it is sufficient to perform
the reduction at 6 numerical kinematic points.
Furthermore, we were able to achieve a computational speedup by
performing all calculations in a finite field with a 10-digit
cardinality, before reconstructing the simple rational numbers
from their finite-field images using Wang's algorithm~\cite{vonManteuffel:2014ixa,Peraro:2016wsq,Wang:1981:PAU:800206.806398}. 

Inserting the symbol of
the master integrals, we directly obtain the symbol of the 
two-loop five-point $\NeqFour$ amplitude.
The amplitude is naturally
decomposed into parity-even and parity-odd parts under a sign-flip 
of `$\tr_5$\!' defined in \eqref{eq:tr5Def}.
At symbol level, the parity grading can be
determined by counting the number of parity-odd letters, $W_{26},...,W_{30}$, in a 
given symbol tensor. The parity-odd part of our result is highly
constrained by the first- and second-entry conditions, as well 
as the integrability of the symbol \cite{Goncharov:2010jf},
leading to a much simpler structure than the even part. 
It is important to note that in all collinear limits the
parity-odd parts of the amplitude vanish since the external 
momenta span only a $3$-dimensional space and hence $\tr_5=0$. 
We attach the explicit symbol-level results for the partial
amplitudes in the ancillary file {\tt amplitudes.m}.
 
\vspace{-8pt}
\section{Validation}\label{sec:validation} 
\vspace{-6pt}
%
%
In the previous section we described the assembly of the two-loop five-point amplitude in $\NeqFour$ in terms of pure master integrals. 
In this section we validate our final result by
checking nontrivial identities between different terms and
verifying universal behavior in kinematic limits. We focus our
discussion on verifying collinear factorization when two external momenta become 
parallel \cite{Bern:2004cz}. Aside from this check, we 
also verified that:
\begin{itemize}
	\item The planar amplitude matches the BDS ansatz \cite{BDS} stating that four- and five-particle amplitudes in planar $\NeqFour$ are given to
		all orders by exponentiating the one-loop amplitude \cite{Bern:1993mq}.
	\item The partial amplitudes satisfy the group-theoretic Edison-Naculich 
		relations \cite{Edison:2011ta}, allowing us to write all subleading single-trace partial
		amplitudes $A^{\text{SLST}}$ in terms of
		linear combinations of planar $A^{\text{ST}}$ and double-trace
		$A^{\text{DT}}$ amplitudes, e.g.
		\begin{align}
		\hspace{.35cm}
		\ASLST{12345}  = &\  5 \AST{13524}  \\
			& \pl \hspace{-.1cm} \sum_{\text{cyclic}}\hspace{-.1cm} \bigg[\AST{12435}\mi 2\AST{12453}  \nonumber \\
			& \hspace{.35cm}\pl \frac{1}{2}\! \left(\ADT{12|345}\mi\ADT{13|245}\right)\!\! \bigg] \,,			\nonumber
		\hspace{-.8cm}
		\end{align}
		\vskip -.3cm
		where the five cyclic permutations are generated by
                 the relabeling $i\to i\pl1$ (mod 5).
                 Thus we need not discuss $A^{\text{SLST}}$ further, and
                 the amplitude is fully specified by two functions,
                 $\text{M}^{\text{BDS}}_{(2)}$ and $g^{\text{DT}}_{234}$.
	\item The infrared poles of the amplitude match the universal pole
		structure predicted by Catani \cite{Catani:1998bh},
		see also e.g.~\cite{Bern:2004cz,Aybat:2006mz}, where
		the poles of two-loop amplitudes can be computed in
		terms of known tree- and one-loop amplitudes.
\end{itemize}
Several of these checks require the one-loop five-point 
amplitudes expanded through order $\eps^2$.
An exact expression for the integrand of this amplitude is known \cite{DimShift}. The box integrals are known to all orders in  
$\eps$ \cite{Bern:1993kr}. The only integral that is not known to all orders is the six-dimensional scalar pentagon, whose symbol can either be computed to
any order in $\eps$ from~\cite{Abreu:2017mtm} or by direct
evaluation of the integral with 
\texttt{HyperInt}~\cite{Panzer:2014caa}. 
We denote by $\I^{d=6\mi2\eps}_5$ the integral normalized by (minus) the
$\tr_5$ of \eqref{eq:tr5Def}, so that it is a pure parity-odd function, and
we give its symbol in the ancillary file {\tt purePentagon6d.m}.

The test we discuss in more detail is the collinear limit of the double-trace partial amplitudes $A^{\text{DT}}$. As already stated, all 
parity-odd contributions of any partial amplitude vanish in this
limit since $\tr_5\!=\!0$.
For concreteness, in the rest of this section we focus on
$\ADT{15|234}$, which in our conventions is symmetric in the 
$(15)$ indices and totally antisymmetric in $(234)$. 
All other double-trace amplitudes are given by simple 
relabelling. 
Scattering amplitudes obey a universal collinear factorization
equation \cite{Bern:1994zx,Bern:2004cz}. Here, we
discuss the five-point limit $2||3$ where two momenta, $k_2$ and $k_3$,
become collinear $k_2 =\tau P\,,\  k_3 = (1\mi\tau)P$ with 
collinear splitting fraction $\tau$. The two-loop amplitude 
factorizes into 
$\sum^{2}_{\ell=0}\text{Split}^{(\ell)}_{23}(\eps) 
\times \A^{(2\mi\ell)}_{4}(\eps)$:
\begin{align}
\label{eq:col_factorization_23}
\hspace{-.4cm}
\A^{(2)}_5\!\!\stackrel{2||3}{\longrightarrow}  \!\!
\raisebox{-20pt}{\includegraphics[scale=.062]{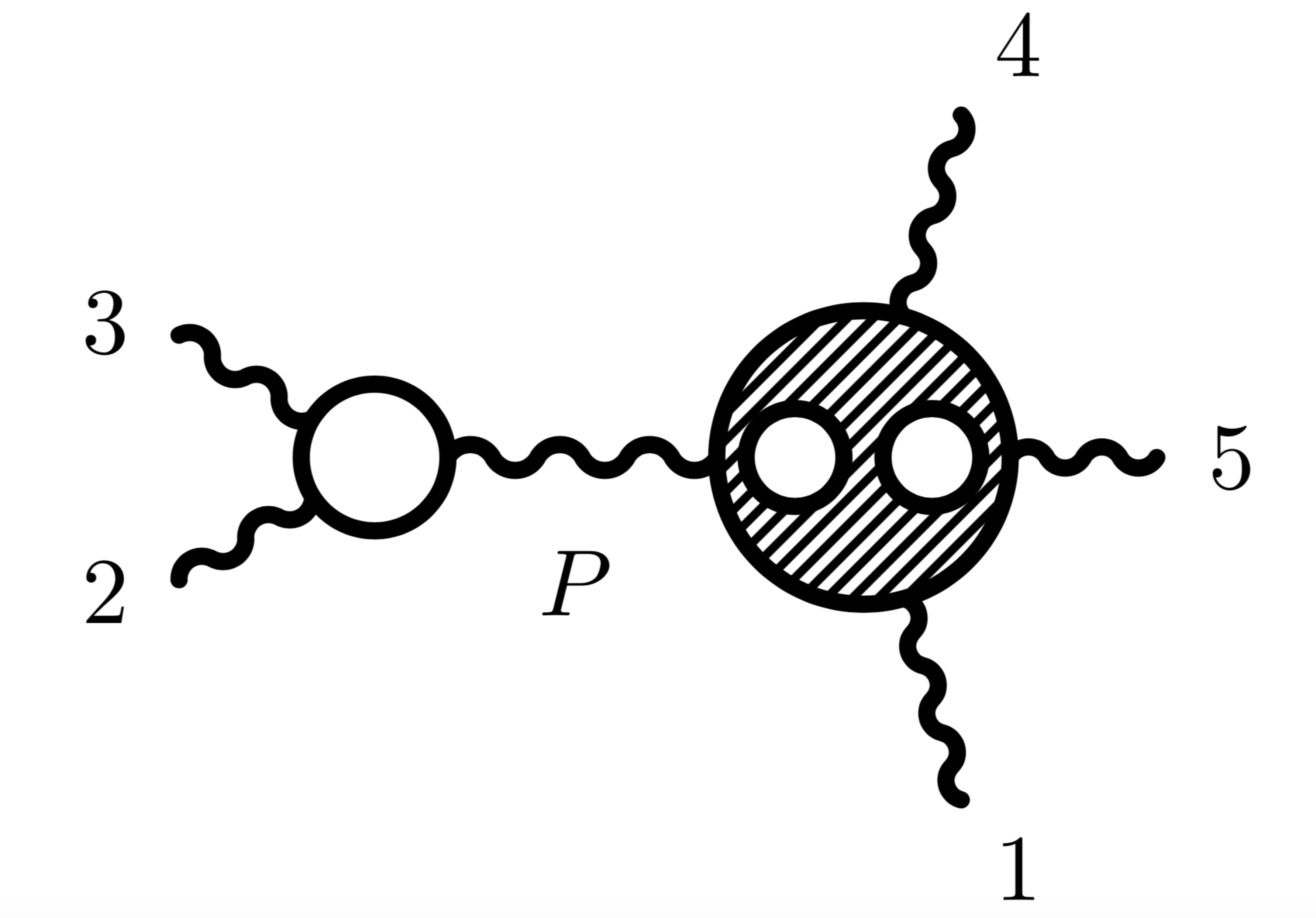}} 
\!\pl\!
\raisebox{-20pt}{\includegraphics[scale=.062]{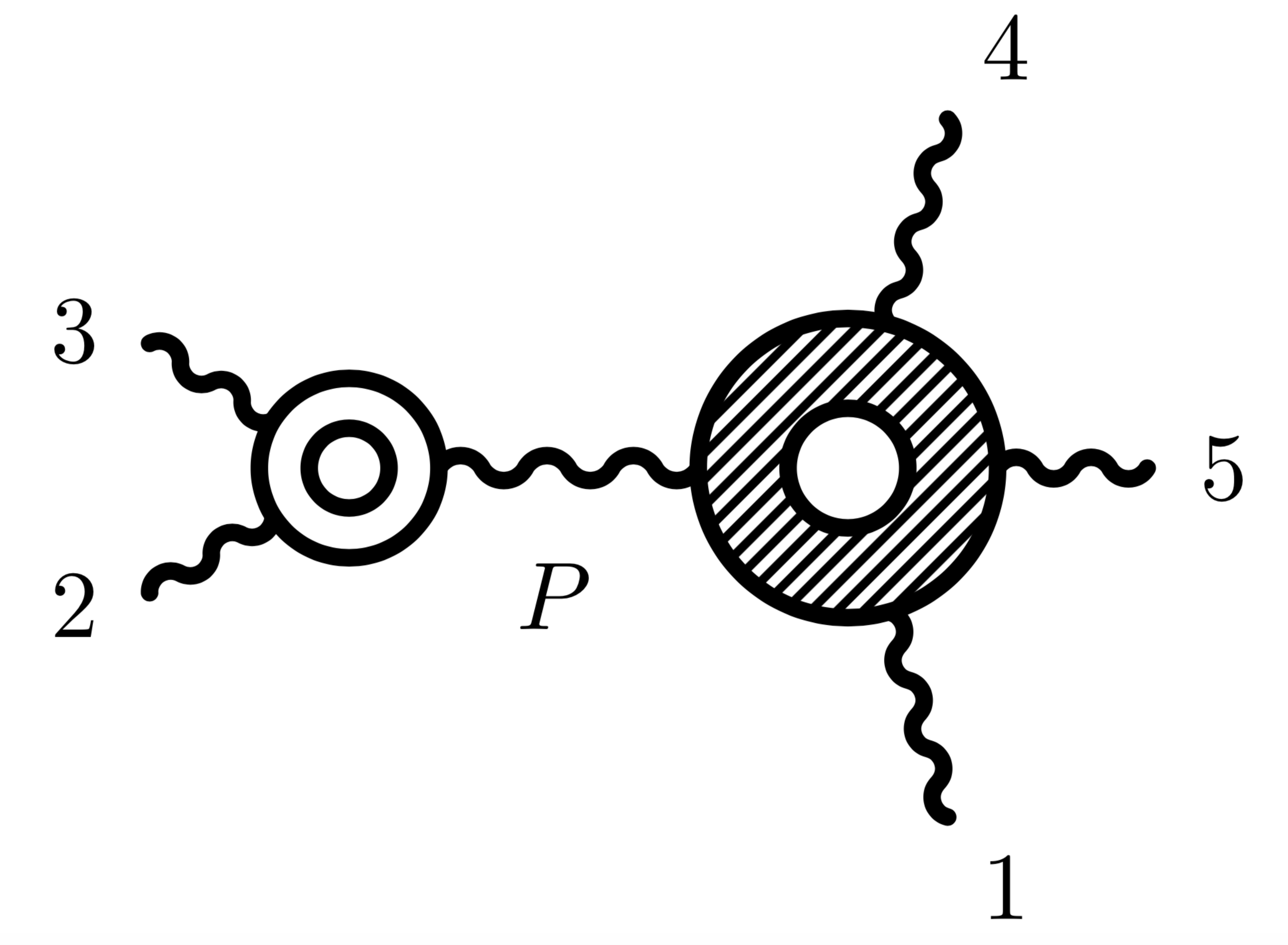}} 
\!\pl\!
\raisebox{-20pt}{\includegraphics[scale=.062]
{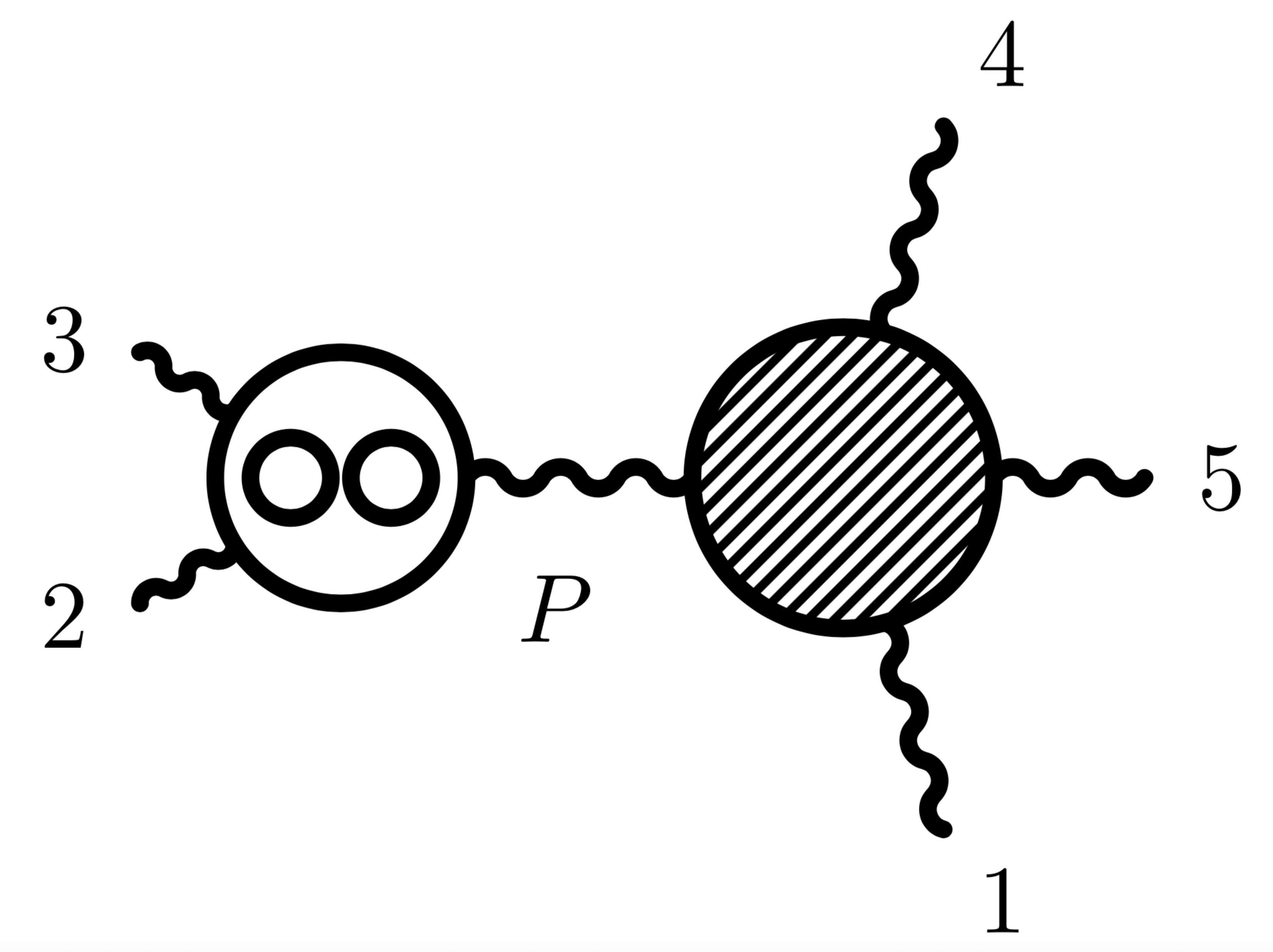}} \,. \hspace{-.3cm}
\end{align}
The empty blobs on the left of each diagram denote the collinear
splitting functions and the filled blobs on the right are the
four-point amplitudes depending only on $P$, $k_1$, $k_4$ and
$k_5$. 
The color part of the splitting function is very simple: in the
example above it is directly proportional to $f^{23P}$.
Kinematic expressions for the one- and two-loop splitting 
functions can be found in 
\cite{Bern:1994zx,Bern:2004cz}.
Furthermore, the one- and two-loop four-point amplitudes 
\cite{Bern:1994zx,BRY}, and relevant integrals 
\cite{Tausk:NPDoubleBox,Smirnov:PlanarDoubleBox}, are also known
to the required order in the $\eps$-expansion. To approach the
collinear limit, we map from the generic five-dimensional
kinematic space (parametrized in terms of the $x_i$ of 
\cite{Badger:2013gxa}) to the collinear limit. This can be done
via the following substitution (see footnote \ref{footnoteKin}):
\eq{
\hspace{.1cm}
x_1\! \mapsto\! s \tau\,, \ 
x_2 \! \mapsto\!  c s \delta \,, \ 
x_3 \! \mapsto\!  r_2c s \delta\,, \ 
x_4 \! \mapsto\!  \delta\,, \ 
x_5 \! \mapsto\!  \mi \frac{1}{c\delta}\,,
\hspace{-.3cm}
}
where $s$ characterizes the overall scale of all Mandelstam 
invariants, $\delta\to0$ corresponds to the collinear limit, 
$\tau$ is the aforementioned collinear splitting fraction, 
$r_2=\frac{\sab{15}}{\sab{45}}$ is the ratio of Mandelstam 
invariants of the underlying four-point process, and 
$c \sim \frac{\sqb{23}}{\ab{23}}$ corresponds to an azimuthal 
phase. Expanding the 31 letter alphabet to leading order in 
$\delta$, we find 14 multiplicatively independent letters in the collinear limit: 7
physical $\{\delta, s, \tau, 1\mi\tau, r_2,1\pl r_2, c\}$ (in
fact this number reduces to 6 at leading power
because $c$ and $ \delta$ always appear in the same
combination, $c \delta^2$) and 7 
spurious letters that cannot be part of the (leading power) limit. When 
comparing the collinear limit of our result to the
factorization formula \eqref{eq:col_factorization_23}, we note
that only
Parke-Taylor factors where
legs $2$ and $3$ are adjacent become 
singular. For instance, while $\PT[12345]\!\mapsto\!\frac{1}{\sqrt{\tau(1\mi\tau)}\ab{23}} \PT[1P45]$,  $\PT[12435]$ has no collinear 
singularity in the $2||3$ limit. We find that our result exactly matches the collinear
factorization formula~(\ref{eq:col_factorization_23}). 
Besides this limit, there are two further
inequivalent collinear limits we can check for $\ADT{15|234}$: 
when $1||5$ and $1||2$. When looking at the color factors of 
the appropriate relabelling of (\ref{eq:col_factorization_23}) 
it becomes clear that neither of them contains 
$\tr[15](\tr[234]\mi\tr[432])$ so $\ADT{15|234}$ is forced to 
be nonsingular in these limits. We have checked that our result
indeed reproduces this behavior.

\vspace{-8pt}
\section{Discussion of the result and outlook}\label{sec:discussion} 
\vspace{-6pt}
%
%
After discussing various consistency checks of our answer 
for the two-loop five-point amplitude in $\NeqFour$, let us
briefly summarize some of its analytic features. First, we
highlight in Tab.~\ref{tab:vanishing_summary} that a number of
terms in the $\eps$-expansion vanish, which is of course 
predicted by the Catani formula. We note that some of the 
two-loop master integrals have weight-two odd terms, but 
this contribution is absent from the amplitude.
\begin{table}[h!]
\centering
\begin{tabular}{|c||c|c|c|c|c|}
\hline
& $1/\epsilon^4 \ w_0$ & $1/\epsilon^3\ w_1$ & $1/\epsilon^2\ w_2$ & $1/\epsilon^1\ w_3$ & $\epsilon^0 \ w_4$  \\[1pt]
\hline
$A^{\text{ST}}_{\text{even}}$ & $\checkmark$ & $\checkmark$ & $\checkmark$ & $\checkmark$ & $\checkmark$\\[1pt]
$A^{\text{ST}}_{\text{odd}}$   & 0 & 0 & 0 & $\checkmark$ & $\checkmark$  \\
\hline
$A^{\text{DT}}_{\text{even}}$ & 0 & $\checkmark$ & $\checkmark$ & $\checkmark$ & $\checkmark$\\[1pt]
$A^{\text{DT}}_{\text{odd}}$   & 0 & 0 & 0 & $\checkmark$ & $\checkmark$  \\
\hline
$A^{\text{SLST}}_{\text{even}}$ & 0 & 0 & $\checkmark$ & $\checkmark$ & $\checkmark$\\[1pt]
$A^{\text{SLST}}_{\text{odd}}$   & 0 & 0 & 0 & 0 & $\checkmark$  \\
\hline
\end{tabular}
\caption{\label{tab:vanishing_summary} Summary of vanishing ($0$)
and non-vanishing ($\checkmark$) terms in the $\eps$-expansion
of the different partial amplitudes.}
\end{table}

We also note that our answers for the amplitude,
as well as individual pure master integrals, are compatible
with the empirical \emph{second-entry-conditions} first 
observed for individual integrals in \cite{Gehrmann:2015bfy,
Chicherin:2017dob,Gehrmann:2018yef,Chicherin:2018mue}. It would 
be very interesting to understand the underlying physical 
reason for this property, perhaps from the point of view of a 
diagrammatic coaction principle \cite{Panzer:2016snt,
Abreu:2017enx,Abreu:2017mtm}.

Our full result is too lengthy to print in this letter. However, it has very restricted analytic structure. For instance, the parity-odd transcendental part of {\it any} derivative of {\it any} weight 4 function in the amplitude belongs to a 12-dimensional subspace of the 111-dimensional space of weight 3 parity-odd functions that obey integrability and the second-entry condition of~\cite{Chicherin:2017dob}.  This 12-dimensional subspace is spanned by the 12 inequivalent permutations, $\Sigma_j$, of the $\O(\eps^0)$ part of the pure, parity-odd scalar pentagon in $d=6$, $\I^{d=6}_5(\Sigma_j)$. (Due to the dihedral $D_5$ invariance of the integral, there are only $5!/10 = 12$ inequivalent permutations.)  The parity-odd part of the $1/\epsilon$ coefficient of $\text{M}^{\text{BDS}}_{(2)}$ is just $-5\I^{d=6}_5(\{12345\})$.

Let us recall that the amplitude is fully specified by $g^{\text{DT}}_{234}$ and the previously-known $\text{M}^{\text{BDS}}_{(2)}$. We may write the odd transcendental part of the derivative of the odd part of $g^{\text{DT}}_{234}$ using this $\I^{d=6}_5$ basis, as
\eq{
\hspace{.4cm}
\partial_{x_i}\! \left[g^{\text{DT,odd}}_{234}\right]\!\Big|_{\text{odd}} \!= \sum_{j,\gamma}   \I^{d=6}_5(\Sigma_j)\ m_{j\gamma}\ \frac{\partial \log W_{\gamma}}{\partial x_i}\,, 
\hspace{-.4cm}
}
where $j$ labels the 12 inequivalent pentagon-permutations 
$\{12543\}$, $\{12453\}$, $\{13524\}$, $\{12534\}$, 
$\{13254\}$, 
$\{12354\}$, 
$\{14325\}$, $\{13425\}$, $\{14235\}$, $\{12435\}$, 
$\{13245\}$, 
$\{12345\},$ and $\gamma\! \in\! \{1,\!...,\!5\}\!\cup\!\{16,\!...,\!20\}\!\cup\!\{31\}$ are the nonzero final entries.  The matrix $m_{j\gamma}$ is
\begin{small}
\begin{align*}
\hskip -.7cm
m_{j\gamma}\!=\!\!\left(\!\!\!
\begin{array}{ccccccccccc}
\mi\frac{17}{4} \!&\! \mi\frac{5}{4} \!&\! \mi6 \!&\! \mi\frac{17}{4} \!&\! \mi\frac{7}{2} \!&\! \mi\frac{17}{4} \!&\! \mi\frac{7}{4} \!&\! \frac{1}{2} \!&\! \mi1 \!&\! \mi\frac{17}{4} \!&\!\! 10 \\[2pt]
 \frac{17}{4} \!&\! \frac{5}{4} \!&\! \frac{5}{4} \!&\! \frac{17}{4} \!&\! 4 \!&\! \frac{17}{4} \!&\! \frac{11}{2} \!&\! \frac{17}{4} \!&\! \frac{1}{2} \!&\! \frac{1}{2} \!&\!\! \mi10 \\[2pt]
 0 \!&\! 0 \!&\!0 \!&\! 0 \!&\! 0 \!&\! 0 \!&\! 0 \!&\! 0 \!&\! 0 \!&\! 0 \!&\! 0 \\[2pt]
 \frac{1}{4} \!&\! \mi\frac{1}{4} \!&\! \mi\frac{1}{4} \!&\! 0 \!&\! 0 \!&\! 0 \!&\! 0 \!&\! 0 \!&\! \frac{1}{4} \!&\! 0 \!&\! 0 \\[2pt]
 0 \!&\! \mi\frac{1}{4} \!&\! \mi\frac{1}{4} \!&\! \frac{1}{4} \!&\! 0 \!&\! 0 \!&\! 0 \!&\! 0 \!&\! 0 \!&\! \frac{1}{4} \!&\! 0 \\[2pt]
 \mi\frac{17}{4} \!&\! \mi6 \!&\! \mi\frac{5}{4} \!&\! \mi\frac{17}{4} \!&\! \mi\frac{7}{2} \!&\! \frac{1}{2} \!&\! \mi\frac{7}{4} \!&\! \mi\frac{17}{4} \!&\! \mi\frac{17}{4} \!&\! \mi1 \!&\!\! 10 \\[2pt]
 \mi\frac{1}{4} \!&\! \frac{1}{2} \!&\! \frac{1}{2} \!&\! \mi\frac{1}{4} \!&\! 0 \!&\! 0 \!&\! 0 \!&\! 0 \!&\! \mi\frac{1}{4} \!&\! \mi\frac{1}{4} \!&\! 0 \\[2pt]
 \frac{1}{4} \!&\! \mi\frac{1}{2} \!&\! \frac{1}{4} \!&\! 0 \!&\! \mi\frac{1}{2} \!&\! \frac{1}{4} \!&\! \mi\frac{1}{4} \!&\! \mi\frac{1}{2} \!&\! 1 \!&\! 0 \!&\! 0 \\[2pt]
 0 \!&\! \frac{1}{4} \!&\! \mi\frac{1}{2} \!&\! \frac{1}{4} \!&\! \mi\frac{1}{2} \!&\! \mi\frac{1}{2} \!&\! \mi\frac{1}{4} \!&\! \frac{1}{4} \!&\! 0 \!&\! 1 \!&\! 0 \\[2pt]
 \mi\frac{1}{4} \!&\! 0 \!&\! \frac{1}{4} \!&\! 0 \!&\! \frac{1}{2} \!&\! \frac{1}{4} \!&\! \frac{1}{4} \!&\! 0 \!&\! \mi\frac{1}{2} \!&\! \mi\frac{1}{2} \!&\! 0 \\[2pt]
 0 \!&\! \frac{1}{4} \!&\! 0 \!&\! \mi\frac{1}{4} \!&\! \frac{1}{2} \!&\! 0 \!&\! \frac{1}{4} \!&\! \frac{1}{4} \!&\! \mi\frac{1}{2} \!&\! \mi\frac{1}{2} \!&\! 0 \\[2pt]
 \frac{17}{4} \!&\! 6 \!&\! 6 \!&\! \frac{17}{4} \!&\! 9 \!&\! \mi\frac{1}{2} \!&\! 4 \!&\! \mi\frac{1}{2} \!&\! \mi\frac{5}{4} \!&\! \mi\frac{5}{4} \!&\!\! \mi10 \\[2pt]
\end{array}
\!\!\!\right)
\hskip -.5cm
\end{align*}
\end{small}
\!\!which has rank 8, so only eight independent
combinations of final entries appear.
For concreteness, we give the symbol of $\I^{d=6}_5(\{12345\})$ in the ancillary
file {\tt purePentagon6d.m}.

While the first derivatives are quite constrained, the second derivatives (actually the $\{2,1,1\}$-coproducts) of the $\eps^0$-terms of the amplitude span the entire $79$-dimensional space identified in~\cite{Chicherin:2017dob}.

Building on this first analytic result for a nonplanar two-loop 
five-point amplitude, there are a number of avenues for future
research. The upcoming work of \cite{LanceToAppear} will explore
the analytic structure of the factorization of the amplitude 
when
one of the external gluons becomes soft. For this limit, there 
exists an eikonal semi-infinite Wilson line picture. Starting 
at two loops the possibility of coupling three hard lines via 
nontrivial color connections opens up, which leads to an 
interesting parity-odd component of the soft-emission function 
which is compatible with the soft limit of our symbol-level 
result. Furthermore, it would be interesting to explore the 
subleading-in-color behavior of this scattering amplitude in 
multi-Regge kinematics \cite{DelDuca:2001gu,Caron-Huot:2017zfo,
DelDuca:2018euv}. With our result, it now also becomes possible 
to test the proposed relation between scattering amplitudes and 
Wilson loops beyond the leading term in the large $N_c$ limit 
\cite{Ben-Israel:2018ckc}, and it would be interesting to match 
our result to a future near-collinear OPE computation on the 
Wilson-loop side.

Since we have now computed the symbol of all relevant Feynman
integrals for massless two-loop five-point scattering, we can 
in principle discuss other theories, such as $\N<4$ sYM as 
well as $\N\geq4$ supergravity. In particular, it would be 
interesting to investigate the uniform transcendentality (UT) 
property of two-loop five-point amplitudes in $\N=8$ supergravity. 
According to \cite{Bourjaily:2018omh}, this integrand only has 
logarithmic singularities and no poles at infinity, so one 
would expect a UT result. Finding such a result would lend
further credence to the empirical relation between logarithmic 
poles of the integrand and transcendentality properties of 
amplitudes.

\vspace{-10pt}
\section{Acknowledgments}
\vspace{-8pt}
L.D.\ and E.H.\ thank Falko Dulat and Hua Xing Zhu for valuable discussions, as well as Huan-Hang Chi and Yang Zhang for initial collaboration on a related project. 
We thank Harald Ita for useful comments on the manuscript.
The work of S.A.~is supported by the Fonds de la Recherche Scientifique--FNRS, Belgium.
The work of L.D.\ and E.H.\ is supported by the U.S. Department of Energy (DOE) under contract DE-AC02-76SF00515.
The work of M.Z.\ is supported by the Swiss National Science Foundation under contract SNF200021 179016 and the European Commission through the ERC grant pertQCD.
The work of B.P. is supported by the French Agence Nationale pour la Recherche, under grant ANR–17–CE31–0001–01.
We thank the Galileo Galilei Institute for Theoretical Physics for hospitality and the INFN for partial support. L.D. acknowledges support by a grant from the Simons Foundation (341344, LA).

\bibliographystyle{apsrev4-1} 
\bibliography{amp_refs} 

\end{document}